\documentclass[aps,pre,twocolumn,superscriptaddress,showpacs]{revtex4-1}
\usepackage{graphicx}
\usepackage{dcolumn}
\usepackage{bm}
\usepackage{epsfig}
\usepackage{color}

\begin{document}


\title{A vehicle with a two-wheel steering system mobile in shallow dense granular media}


  
\author{Po-Yi Lee}
\affiliation{Department of Mechanical Engineering, National Taiwan University, Taipei 106, Taiwan}
\author{Meng-Chi Tsai}
\affiliation{Department of Mechanical Engineering, National Taiwan University, Taipei 106, Taiwan}
\author{I-Ta Hsieh}
\affiliation{Department of Mechanical Engineering, National Taiwan University, Taipei 106, Taiwan}
\author{Pin-Ju Tseng}
\affiliation{Department of Mechanical Engineering, National Taiwan University, Taipei 106, Taiwan}
\author{Guo-Jie Jason Gao}
\email{koh.kokketsu@shizuoka.ac.jp, gjjgao@gmail.com}
\affiliation{Department of Mechanical Engineering, National Taiwan University, Taipei 106, Taiwan}
\affiliation{Department of Mathematical and Systems Engineering, Shizuoka University, Hamamatsu, Shizuoka 432-8561, Japan}

\date{\today}

\begin{abstract}
We design a vehicle with a steering system made of two independently rotatable wheels on the front. We quantify the effectiveness of the steering system in the mobility and maneuverability of the vehicle running in a box containing a layer ping-pong balls with a packing density $0.8$, below the random close packing value $0.84$ in 2D. The steering system can reduce the resistance exerted by the jammed balls formed ahead of the fast-moving vehicle. Moreover, if only one of the two steering wheels rotates, the vehicle can turn into the direction opposite to the rotating wheel. The steering system performs more efficiently if the wheels engage the ping-pong balls better by increasing the contact area between the wheels and the balls. We advocate applying our design to machines moving in granular materials with a moderate packing density.
\end{abstract}


\maketitle

\section{Introduction}
\label{introduction}

There has been a long history of designing mobile machines, for example airplanes and ships, in continuously deformable thermal media including air and water. However, similar attempts in deformable athermal media such as granular materials are only very recent. Partly it is because of the complexity of designing a mobile machine within a medium with equal stiffness. More importantly, the nonequilibrium and discrete features of athermal media make formulating a universal governing equation describing their rheology a very difficult task \cite{behringer96}, except under specific conditions approximating the media as a continuum \cite{tsimring02, goldman13, goddard14, kamrin17}. In the literature, most designs for a vehicle in granular materials were initiated by mimicking creatures in nature, including sandfish and lizard \cite{herrmann09, goldman09_2, goldman13, goldman14}, insects \cite{goldman09_1} and clam \cite{hosoi14, goldman15}. Similar efforts using artificial designs are only lately \cite{goldman11}. Compared with designs inspired by living creatures such as a clam that uses an intricate way to move forward by swallowing and discharging sands, artificial ones have to meet the challenges of being simpler and more maneuverable. They should perform more precise movements from moving along a straight line to turning at a sharp angle. On the contrary, a sandfish can only move in a more or less sinuous way. These challenges make designing artificial vehicles a demanding job, but at the same time they are better candidates to satisfy more practical applications, for example, a robotic rover used on a low gravity extraterrestrial planet covered by sand that traps it easily. 

Recent studies on horizontally pulling an object in dense granular materials revealed that there is a highly jammed region formed ahead of the fast-moving object, where force chains between jammed grains frequently build and destruct \cite{okumura14, gondret16, ito16, hayakawa17}. Unlike a vehicle in thermal fluids, which usually has its steering system located at the end, in this study we design a vehicle in athermal granular materials, equipped with a steering system placed at its front by utilizing this jammed region. The steering system is composed of two independently rotatable wheels which can interact with the jammed region to reduce the resistance experienced by the vehicle and allow the vehicle to change its moving direction freely in a dense granular medium.

We test our vehicle in a pool containing a layer of moderately packed ping-pong balls and quantitatively measure its performance in several vehicle setups. Our experimental results show clear evidence of the existence of the reported jammed region, formed in front of the fast-moving vehicle. The steering system can mobilize the jammed region to allow the vehicle to move faster or escape from being trapped by the jammed balls. Moreover, if only one of the two steering wheels rotates, the vehicle can turn to the direction away from the rotating wheel. The turning mechanism becomes more effective if the rotating wheel engages the ping-pong balls better by increasing the contact area with the balls.

Below we introduce the design of the vehicle and the two-wheel steering system in section \ref{experimental}, and show our experimental results of operating the vehicle in a shallow ping-pong ball pool in section \ref{results}. We conclude our study in section \ref{conclusions}.

\section{Experimental Setup}
\label{experimental}

\begin{figure}
\vspace*{0.25in}
\includegraphics[width=0.45\textwidth]{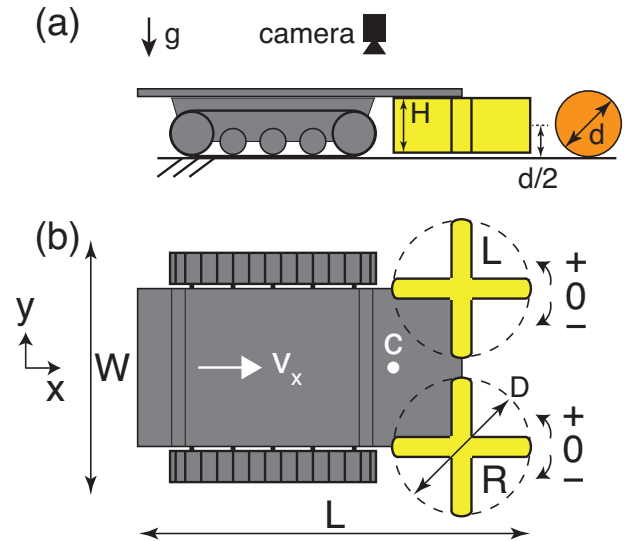}
\caption{\label{fig:schematic} (color online) Schematic of the vehicle with left (L) and right (R) steering wheels (yellow objects), which can be still (0), rotate counterclockwise (+) or clockwise (-) separately. A mark on the vehicle ($C$) is used to identify its position, $(x,y)$, on the horizontal $xy$-plane. The arrow $g$ shows the direction of gravity. (a) side view. (b) top view.}
\end{figure}

We use ready-made parts manufactured by Tamiya, a Japanese model company. The vehicle is composed of a moving unit (Tamiya item No. 70104) and a two-wheel steering system, located in front of the moving unit. The moving unit, which drives the vehicle forward or backward, contains a twin-motor gearbox (Tamiya item No. 70097), equipped with two FA-130 regular motors powered together by two 1.2 volts, 2,450 mAh Ni-MH batteries. The twin-motor gearbox drives two independent but otherwise identical caterpillar tracks and allows certain differential during turning. The steering system contains two separate gearboxes (Tamiya item No. 70103), each equipped with a FA-130 regular motor powered by two 1.2 volts, 2,450 mAh Ni-MH batteries and driving a rotatable plastic steering wheel in the steering system. Each steering wheel basically has a cross-shape, or can be replaced by a round-shaped one to test the shape effect. The above design guarantees that no battery supplies electricity to more than one action of the vehicle, which maintains the consistency and controllability of the experiments. Besides, we make sure that all batteries are drained by no more than $3\%$ of the fully-charged voltage in all experiments. Therefore, comparing results from experiments with different motor performance due to batteries drained variously is not a concern. All gearboxes are set at their highest gear ratios to output the maximum available torque. A schematic of the vehicle is shown in Fig. \ref{fig:schematic}.

The vehicle, having approximate dimensions of $11\ cm$ in width ($W$) and $23\ cm$ in length ($L$), can move freely in a wood box of $63\ cm$ ($W_b$) by $108\ cm$ ($L_b$), which contains a layer of $N$ loosely packed plastic ping-pong balls with an average diameter of $3.8\ cm$ $(d)$. The area packing density $\phi$ of the system is defined as
\begin{equation} \label{spherical_approx0}
\phi  = [N\pi {(d/2)^2} + WL]/{W_b}{L_b}.
\end{equation}
The two steering wheels measure about $2.8\ cm$ high ($H$) and $6.0\ cm$ in diameter ($D$). Left and right wheels viewed from above the vehicle are labeled by L and R, respectively. Their vertical geometric centers are located at $d/2$ above the bottom of the box. Each of them can be either still (status: 0), or rotate counterclockwise (status: +) or clockwise (status: -). The status of the steering system can be expressed as (L: status / R: status). The time-lapse positions of the vehicle, $(x,y)$, is captured by a digital camera (Logitech C-920r) with a top-down view of the box. The optical distortion of the recorded area introduced by the camera is negligible. We calculate the average velocity $v_x$ of the vehicle using $v_x=\Delta x/\Delta t$. In the following analysis, $\Delta x=x_2- x_1=10\ cm$ and $\Delta t = t_2 - t_1\ sec$, calculated by obtaining the time difference between two still images, 1 and 2, taken at $t=t_1$ and $t=t_2$, respectively.

\begin{figure}
\includegraphics[width=0.45\textwidth]{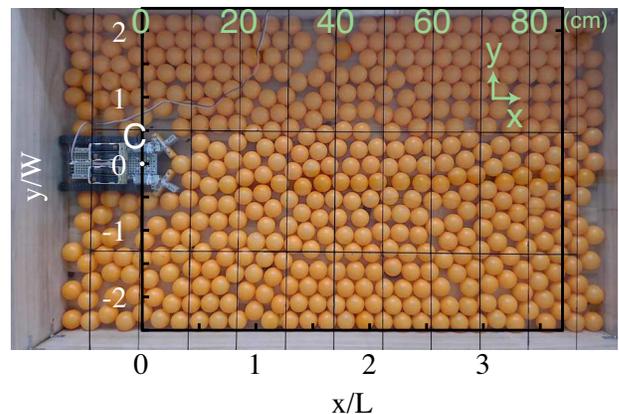}
\caption{\label{fig:IC} (color online) The initial condition of an experiment, $N=460$ and $\phi = 0.8$. The horizontal-black and vertical-white tick labels associated with the black frame show normalized positions, $x/L$ and $y/W$, of the vehicle, as used in Fig. \ref{fig:comparison_vx}, \ref{fig:comparison_y} and \ref{fig:comparison_y_wheel_shape}. The horizontal green labels show the actual $x$ location in $cm$ in the box, measured from the initial position of the vehicle.}
\end{figure}

To take the averages and their error bars of the position and velocity of the vehicle, we repeat one experiment with a given set of parameter three times, and the results are independently verified by all authors, who separately build their own vehicles using the same design. We observe the same results qualitatively in all experiments executed by different authors. We prepare a layer of ping-pong balls in the box at a required packing fraction $\phi$ by randomly pouring balls into the box and waiting until all balls stop moving. Although there are some regions where balls are packed orderly, the whole packing is overall disordered as long as the area packing fraction $\phi$ remains reasonably smaller than the random close packing density, $0.84$, in two dimensional (2D). There is enough space above balls so that they can pile on top of one another if needed and therefore the system is quasi-2D. A snapshot of an initial condition is shown in Fig. \ref{fig:IC}. In the following section we present and discuss the findings extracted from experiments conducted by one representative person.

\section{Results and Discussion}
\label{results}

Here we describe the results of our experiments on the vehicle with a two-wheel steering system, introduced in section \ref{experimental}. The vehicle moves in a shallow granular medium of plastic ping-pong balls with a given area packing density, $\phi$. Initially, we examine the mobility and maneuverability of the vehicle moving at three values of $\phi$, $0.6$, $0.7$ and $0.8$ (results not shown). No substantial jammed region forms in front of the vehicle at the two lower densities of $0.6$ and $0.7$, and therefore the two-wheel steering system performs poorly in these two situations. As a result, we decide to use $\phi=0.8$, which best demonstrates the effectiveness of the steering system, throughout this study. The following experiments are all performed with the same initial condition, $N=460$ ping-pong balls and $\phi = 0.8$.

\begin{figure}
\vspace*{0.25in}
\includegraphics[width=0.45\textwidth]{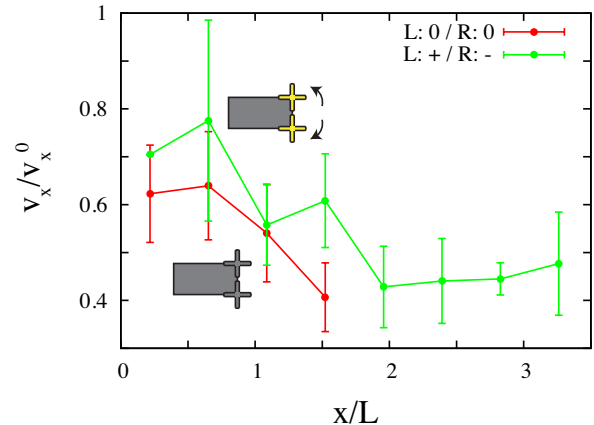}
\caption{\label{fig:comparison_vx} (color online) Average normalized velocity $v_x/v_x^0$ of the vehicle with the two steering wheels set at (L: 0 / R: 0) (red line, terminated at $x/L\approx1.5$, because the vehicle could occasionally be fully blocked by jammed balls) or (L: + / R: -) (green line). $v_x^0$ is the speed of the vehicle in an empty box. $\phi = 0.8$.}
\end{figure}

To demonstrate that ping-pong balls can jam ahead of the moving vehicle with still steering wheels (L: 0 / R: 0), which slows down or even blocks the motion of the vehicle, we measure its normalized velocity in the $x$-direction, $v_x/v_x^0$, where $v_x^0$ is the velocity of the vehicle moving in the box containing zero ping-pong balls. Then we turn on the steering-system (L: + / R: -), and the two wheels rotate in opposite directions and symmetrically push jammed balls ahead of the vehicle to its sides. Another setting, (L: - / R: +), where the two wheels swipe jammed balls toward the center of the vehicle does not work, because eventually balls get into the space between the two steering wheels and block their rotation. The results are shown in Fig. \ref{fig:comparison_vx}. We can clearly see that, in the presence of balls and still steering wheels, the speed of the vehicle in the beginning decreases to about $60\%$ of the speed value in an empty box. The vehicle slows down further as it moves forward and can occasionally reaches a full stop by jammed ping-pong balls in the way. On the other hand, with the steering system rotating, the vehicle moves faster by about $10\%$ and never gets trapped by jammed balls until it reaches the other side of the box.

\begin{figure}
\vspace*{0.25in}
\includegraphics[width=0.45\textwidth]{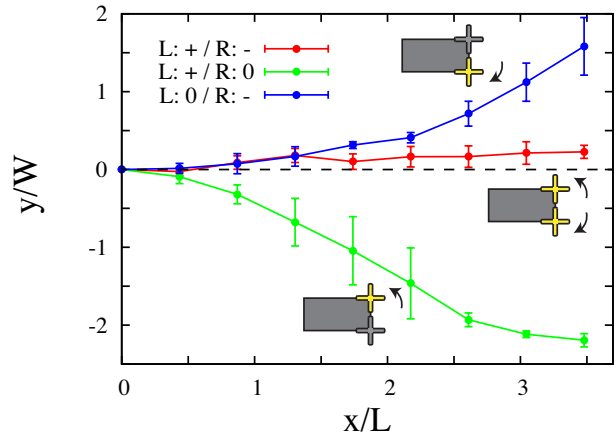}
\caption{\label{fig:comparison_y} (color online) Average trajectories of the vehicle with the two steering wheels set at (L: + / R: -) (red line), (L: + / R: 0) (green line) or (L: 0 / R: -) (blue line). $y/W=0$ (black dashed line) is plotted for eye guidance. $\phi = 0.8$.}
\end{figure}

Then we test if the vehicle can turn if only one steering wheel rotates and the other stays still. We find that the vehicle turns to the direction away from the rotating wheel. The scenario stays the same while both wheels rotate in the same direction, but there is no significant change in the turning efficiency. The results are presented in Fig. \ref{fig:comparison_y}. As expected, a symmetric rotation setting of the steering wheels, (L: + / R: -), allows the vehicle to move alone an almost straight path, while an asymmetric setting, (L: + / R: 0) and (L: 0 / R: -), deviates the vehicle from the straight course. Settings (L: + / R: +) and (L: + / R: 0) produce almost the same effect, so do settings (L: 0 / R: -) and (L: - / R: -). In all experiments, the vehicle never fails to perform a turn with the attempted setting. We can observe that the curved courses in Fig. \ref{fig:comparison_y} do not mirror each other perfectly across the horizontal axis. This is because it is difficult to build a vehicle with a perfect symmetric steering system. This asymmetry can be minimized if we average the trajectories using more trials.

\begin{figure}
\vspace*{0.25in}
\includegraphics[width=0.45\textwidth]{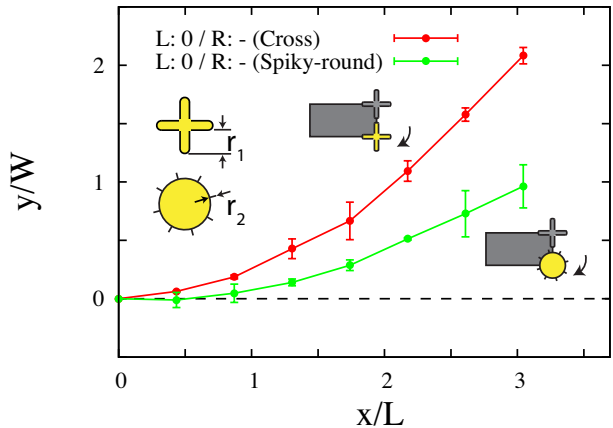}
\caption{\label{fig:comparison_y_wheel_shape} (color online) Average trajectories of the vehicle with the two steering wheels set at (L: 0 / R: -), with a cross-shaped (red line) or spiky round-shaped (green line) rotating wheel. $y/W=0$ (black dashed line) is plotted for eye guidance. $\phi = 0.8$. $d:r_1:r_2=1:0.66:0.11$, where $r_1$ and $r_2$ are the arm length of the cross-shaped wheel and the depth of a spike on the round-shaped wheels, respectively.}
\end{figure}

Finally, to understand the influence of the shape of the steering wheels, we replace one of the cross-shaped wheels with a round-shaped one with a spiky rubber surface, while keeping all other characteristics of the vehicle untouched. The cross-shaped steering wheel and the round-shaped one have similar dimensions and weights. The length of an arm of the cross-shaped wheel and the depth of a spike on the round-shaped wheel are $0.66d$ and $0.11d$, respectively. The results are shown in Fig. \ref{fig:comparison_y_wheel_shape}. It can be obviously seen that the cross-shaped steering wheel works almost twice as effective as the round one in terms of turning the vehicle, because its shape endows larger contact area with the jammed ping-pong balls and exchange momentum with them more efficiently.

\begin{figure}
\vspace*{0.25in}
\includegraphics[width=0.38\textwidth]{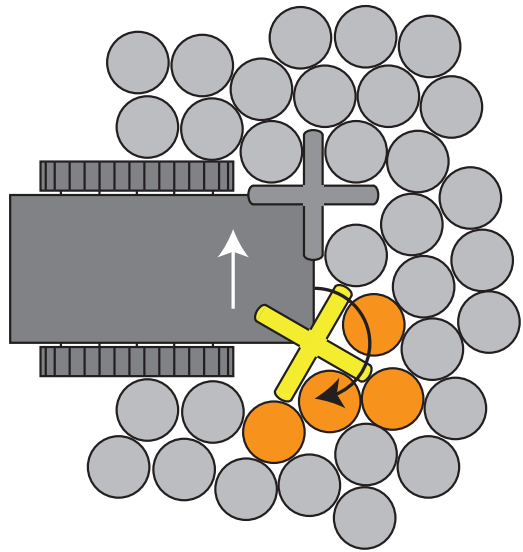}
\caption{\label{fig:mechanism} (color online) Turning mechanism of the vehicle when one of its steering wheel (yellow) rotates. The rotating wheel mobilizes some of its nearby jammed balls (orange), while other balls (grey) stay undisturbed.}
\end{figure}

We can explain the above results by considering the momentum exchange between the moving vehicle with a rotating steering wheel and the jammed ping-pong balls in front of it, as depicted schematically in Fig. \ref{fig:mechanism}. The rotating wheel installed on the vehicle liquidizes its neighboring jammed balls and transfers momentum into them. In exchange, the vehicle gains opposite momentum according to Newton's third law of motion and turns away from the rotating wheel. It is worth noting that this mechanism works only when the vehicle moves fast enough in a fairly packed sea of balls, so that a jammed region can form ahead of the moving vehicle, as described before. The two-wheel steering system becomes ineffective if the vehicle travels too slow or in a loosely packed granular medium.

In this study, we do not adopt a single steering wheel design, because it renders the vehicle a forward-triangle shape, and therefore the vehicle plunges easily into ping-pong balls of smaller size than the steering wheel. Superior to the single-wheel design, the two-wheel steering system offers a larger area to engage balls and makes the vehicle more controllable. Besides, if only one of the steering wheels rotates, the vehicle can turn into the direction opposite to the rotating steering wheel, while the other static one maintains a jammed region of balls ahead of the vehicle, which makes the turning mechanism more stable.

Although we achieve consistent results of controlling the vehicle, there are still several issues should be taken into account. First, the system size is still very small and therefore the finite size effect and the influence from boundaries cannot be ignored. For example, the ping-pong balls in the box ahead of the vehicle can be fully jammed when the vehicle moves beyond $x/L > 2$. Second, we need a well-defined way to prepare an initial condition of randomly placed balls with a given $\phi$. Using a map of computer-generated positions may be a solution to this issue. Finally, it is important to use much smaller balls than the sizes of the steering wheels and the vehicle so that the shape of the jammed region in front of the vehicle can be more flexible and avoid artificial crystallization. We expect the jammed region becomes smaller with decreasing the particle size of the granular media. We will address these issues in our next study.

\section{Conclusions}
\label{conclusions}

In summary, a region formed by jammed ping-pong balls ahead of a fast-moving vehicle is now believed to play a crucial role in affecting the motion of the vehicle. We take advantage of this jammed region and design a vehicle with a front two-wheel steering system, a feature different from most mobile machines used in thermal fluids, where rear steering systems are more common. The proposed steering system not only reduces the resistance experienced by the moving vehicle, but also helps it change directions freely. The turning capability of the vehicle can be improved if the steering wheels interact with the jammed balls more effectively by increasing the contact area between the steering wheels and the balls.

The results of our study have many practical applications. For example, a space rover on another planet, offering very low gravity and covered by sand. Compared with the regular gravity on Earth, the lower gravity makes the sand loosely packed and maybe flow more easily, which cause the rover to sink into the sand more frequently. In order to move smoothly, the rover has to overcome the resistance from the sand jammed and piled up ahead of it, a scenario closely resembles what we have investigated in this study. We believe a similar maneuvering mechanism can be applied to designing a wide range of mobile machines used in dense athermal media, a field still in its infancy.

\section{acknowledgments}
GJG gratefully acknowledges financial support from National Taiwan University funding 104R7417, MOST Grant No. 104-2218-E-002-019 (Taiwan), and startup funding from Shizuoka University (Japan).

\bibliography{paper3}

\end{document}